# Quantum computing for energy systems optimization: Challenges and opportunities

Akshay Ajagekar, Fengqi You[*]

Cornell University, Ithaca, New York 14853, USA

## Abstract

The purpose of this paper is to explore the applications of quantum computing to energy systems optimization problems and discuss some of the challenges faced by quantum computers with techniques to overcome them. The basic concepts underlying quantum computation and their distinctive characteristics in comparison to their classical counterparts are also discussed. Along with different hardware architecture description of two commercially available quantum systems, an example making use of open-source software tools is provided as a first step for diving into the new realm of programming quantum computers for solving systems optimization problems. The trade-off between qualities of these two quantum architectures is also discussed. Complex nature of energy systems due to their structure and large number of design and operational constraints make energy systems optimization a hard problem for most available algorithms. Problems like facility location allocation for energy systems infrastructure development, unit commitment of electric power systems operations, and heat exchanger network synthesis which fall under the category of energy systems optimization are solved using both classical algorithms implemented on conventional CPU based computer and quantum algorithm realized on quantum computing hardware. Their designs, implementation and results are stated. Additionally, this paper describes the limitations of state-of-the-art quantum computers and their great potential to impact the field of energy systems optimization.

[*]Corresponding author. Phone: (607) 255-1162; Fax: (607) 255-9166; E-mail: fengqi.you@cornell.edu

*Graphical abstract*

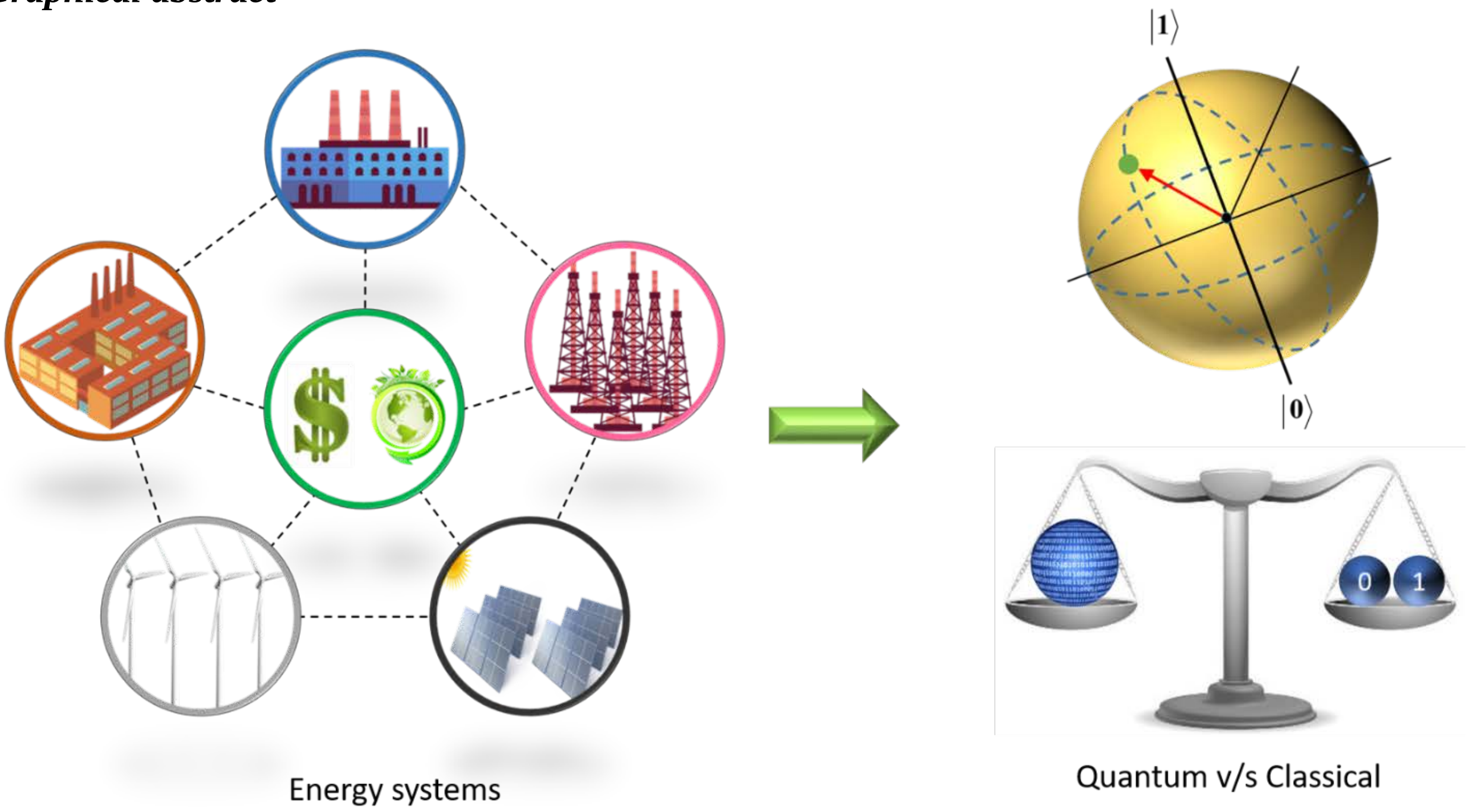

***Keywords:*** Quantum computing, energy systems, optimization

***List of Abbreviations:*** CPU, Central Processing Unit; GPU, Graphics Processing Unit; HENS, Heat Exchanger Network Synthesis; QAOA, Quantum Approximate Optimization Algorithm; QPU, Quantum Processing Unit; QUBO, Quadratic Unconstrained Binary Optimization; UC, Unit Commitment; VQE, Variational Quantum Eigensolver

***Nomenclature***

$A_i$ Zeroth order cost coefficient for unit $i$

$A$, $B$ Weight parameters for QUBO

$B_i$ First order cost coefficient for unit $i$

$C_i$ Second order cost coefficient for unit $i$

$C_{ij}$ Cost of transporting one unit of energy from location $i$ to location $j$

$c_{ij}$ Cost of heat exchanger between source $i$ and sink $j$

$D_j$ Heat demand at sink $j$

| | |
|---|---|
| $f_i$ | Fuel cost of committed unit *i* |
| $h_i$ | Size of grid for power generated by unit *i* |
| $L$ | Load requirement |
| $p_i$ | Power generated by unit *i* |
| $P_{min,i}$ | Minimum possible power generated by unit *i* provided unit is online |
| $P_{max,i}$ | Maximum possible power generated by unit *i* provided unit is online |
| $q_{ij}$ | Heat transfer between source *i* and sink *j* |
| $S_i$ | Heat supply for source *i* |
| $T_{pq}$ | Number of units of energy transported from plant *p* to plant *q* |
| $U$ | Set of units |
| $U_{ij}$ | Maximum possible heat transfer between source *i* and sink *j* |
| $v_i$ | 0-1 variable representing whether unit *i* is offline |
| $w_{ij}$ | 0-1 variable for heat match between source *i* and sink *j* |
| $x_{pi}$ | 0-1 variable for assignment of plant *p* to location *i* |
| $y_i$ | 0-1 variable representing whether unit *i* is online |
| $z$ | 0-1 variable denoting grid points |
| $Z$ | Pauli operator assuming values in {-1, 1} |

# 1. Introduction

With the rising demand for energy and the need for environmental protection, there has been a primary interest in the design, control, and planning of energy systems. New energy resources are also being integrated into energy systems rendering optimal management and control of available resources to be a key issue in harnessing new technologies. Without optimal utilization of resources, the cost of investment for these technologies cannot be justified. Therefore, optimization tools and algorithms provide a suitable way to solve complex energy systems problems in this field. For the comparison between renewable energy sources, factors such as price of generated energy, greenhouse gas emissions, availability of resources, efficiency of energy conversion, natural resource requirement, and social impacts have been previously taken into account [1]. Proper allocation of available energy sources requires the study of models such as energy planning models, energy supply-demand models, forecasting models, renewable energy models, emission reduction models, and optimization models [2]. Energy generation planning and scheduling systems, location and transportation problems, resource allocation, engineering design, network planning are some of other areas wherein optimization methods has found widespread applications [3]. Novel optimization frameworks with comprehensive energy conversion modelling and complete system optimization is an important tool to reduce life-cycle cost and evaluate optimal design for micro-grid energy systems [4]. The worldwide energy crisis demands development of hybrid renewable energy systems instead of fossil fuels. Simulation based studies of large-scale global optimization for such hybrid energy systems through energy management strategies have also been conducted [5]. Reliability and cost of power systems with integrated renewable resources are important aspects considered by system operators. Optimization methods for maintenance by considering the system reliability can be applied for such systems [6]. Apart from the deterministic optimization methods, in the last few decades, heuristic approaches and artificial neural networks have also been proposed. Meta-heuristic algorithms like simulated annealing [7], tabu search [8], genetic algorithms [9], ant-colony optimization [10] have grown quite popular due to their ability to tackle complex optimization problems where traditional methods prove ineffective. As the complexity of some problems increase, exponential computation time might be needed to find the optimum in the worst case. For example, in the case of power grid networks, complexity of the network is a function of the number of power plants that are a part of the grid. According to the data obtained from U.S. energy information administration (EIA),

Fig. 1 shows a continuous rise in the number of power plants in the United States for the last decade, thus complexity of power networks is expected to keep increasing. Hence, there arises a need for new tools and techniques which are capable of addressing such energy systems optimization problems with high complexity, and which provide good solutions in reasonable runtimes.

Quantum computing provides a novel approach to help solve some of the most complex problems while offering an essential speed advantage over conventional computers [11]. This is evident from the quantum algorithm proposed by Shor for factorization which is exponentially faster than any known classical algorithm [12], and Grover's quantum search algorithm capable of searching a large database in time square root of its size [13]. Quantum chemistry [14], machine learning [15], cryptography [16] and optimization are some of the areas where a quantum advantage is perceived when facilitated by a quantum computer. Recent advances in hardware technology and quantum algorithms allow for complex energy systems optimization problems to be solved on a quantum computer. Methods such as process integration, superstructure optimization, and their applications to sustainable design and synthesis of energy systems are crucial and have been an active research area [17]. These methods due to their high complexity require classical optimization solvers to solve the formulated problem, which can be computationally expensive with no guarantee of returning a solution. Classical optimization methods applied to large-scale renewable and sustainable energy systems to perform multi-objective optimization result in use of high computational effort [18]. Physical modelling and optimization of hybrid energy power systems require exponential computation time as the size of problem increases [19]. It is imperative to account for optimal design and multi-scale decisions of shale gas energy systems that are economical, sustainable and socially responsible. Modelling and optimization of such shale gas energy systems under multiple types of uncertainties is a computational challenge [20]. Decomposition of complex large-scale design and synthesis optimization of energy system problems can result in nested optimization problems simpler than the original but is much more computationally intensive [21]. An integral part of designing energy supply chains with multi-scale complexities is to build a large-scale integrated model that includes minute details across all spatial and temporal scales. Classical general purpose optimization methods and off-the-shelf optimization solvers fail to compute feasible solutions for the resulting large-scale multi-objective optimization problem [22]. Optimizing the bioenergy and biofuel supply chain network structure comprising of multiple sites and multiple echelons is performed at the design stage through superstructure optimization. However, a comprehensive and detailed superstructure although appealing, may be computationally intractable [23]. With the increasing importance of water-energy-food nexus, challenges that account for its multiple scales, appropriate system boundaries, modelling the decision making and conflicting objectives along with uncertainties must be considered. Although a higher level of detail provides more thorough details, the associated modelling and computational challenges increase [24]. Custom algorithms for a specific problem class can sometimes outperform the best available classical solvers, but a generic solution approach is much more desirable. While solving large instances of complex problems with deterministic technique is intractable, approximate algorithms should be considered. Quantum computers realize such approximate algorithms intrinsically.

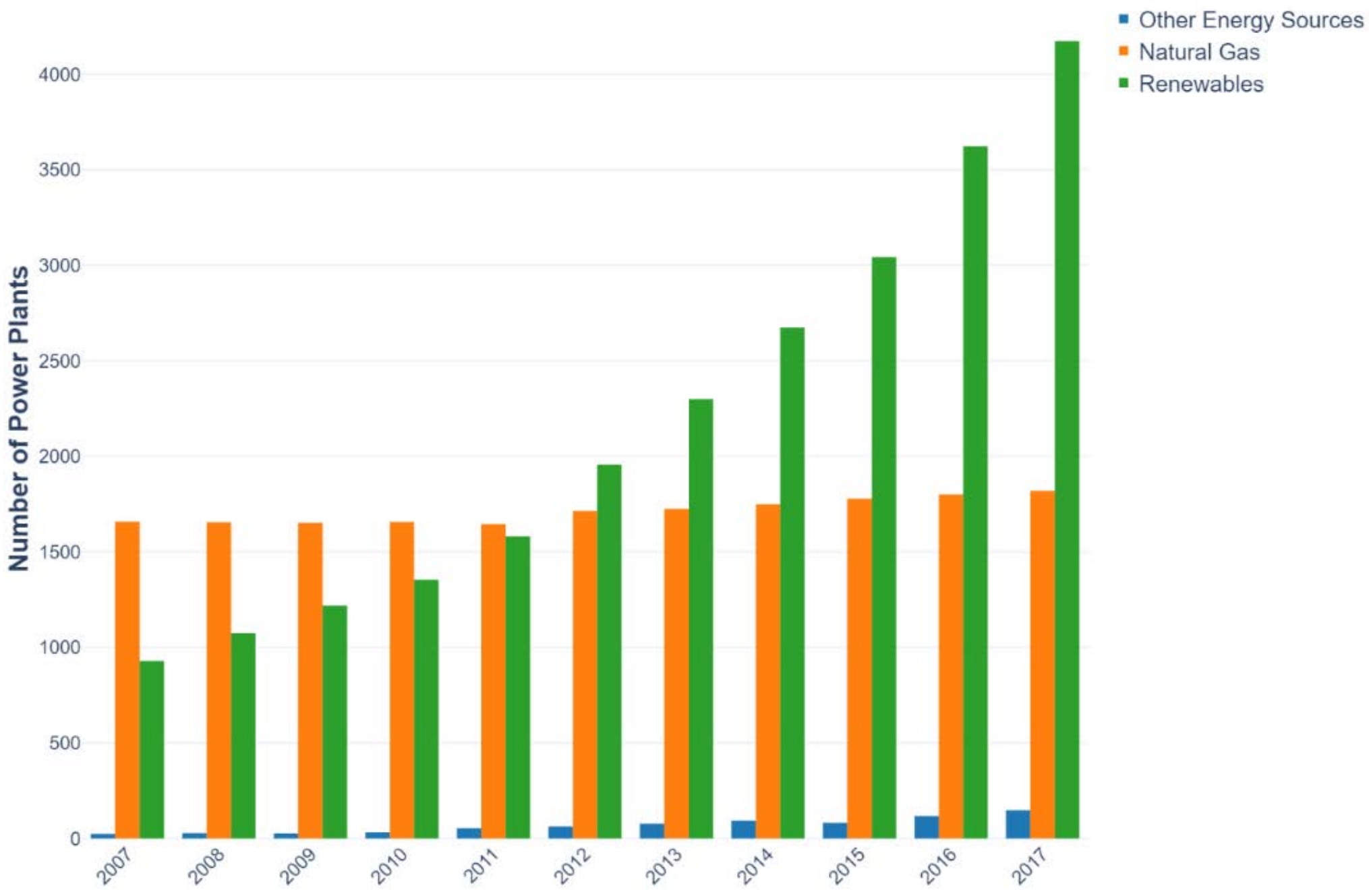

*Fig. 1. Number of power plants in U.S. in the last 11 years*

Important large-scale problems of practical relevance can consist of thousands of variables, and constraints which due to their complex combinatorial nature may require days or even weeks to converge to an optimal solution. Design of shale-gas supply chain network covering more than 10,000 $km^2$ area requires optimizing a mixed-integer linear problem with 51,133 variables and 51,880 constraints. State-of-the-art classical CPU-based solver takes more than 15 hours to compute a solution with the desired optimality gap [25]. Another such example of an integrated optimization framework for energy supply chain involves optimization of a quadratic constrained mixed-integer problem with 8,321 total variables and 7,247 constraints. Solving this problem by a classical CPU-based solver requires 15.6 hours of computation time [26]. Although energy systems problems have been well acknowledged by industry and academia, most of the existing studies are limited to regional scales. Extending such studies to national or global scales faces the limitation of computational power and can result in computationally intractable mathematical problems. All these challenges motivate the need of developing novel and more efficient solution strategies for energy systems optimization problems. Quantum computing is a game-changer technology that is able to tackle such intractable problems and may produce good solutions in reasonable runtimes. Unlike classical computers exploring the entire space of feasible solutions, quantum computers focus on exploring selected feasible subspaces, thus inducing a quantum speedup. This paper exemplifies the promise of quantum computation through well-known energy systems optimization problems.

The purpose of this paper is to introduce readers to the new and emerging field of quantum computation for optimization, and state its implications and applications to energy systems engineering. A brief background unraveling the working of a quantum computer and its major dissimilarities with conventional CPU/GPU-based classical computers will also be provided. Conventional or classical computers are governed by classical physics, and simply put are integrated multi-purpose computers used by most people on a daily basis. Furthermore, examples solving small optimization problems on different types of quantum computers will be illustrated here allowing the readers to get a basic idea on using quantum computers for optimization purposes. Finally, a detailed methodology for application of quantum computers to classes of energy systems optimization like facility location allocation, unit commitment problems and heat exchanger network synthesis, is discussed. Some of the challenges faced by quantum computing for energy systems optimization, as well as its limitations and scope in the future, are also discussed. The major contributions of this paper are listed as follows:

- A brief review of quantum computing fundamentals and operating functionality of two types of quantum computing architectures has been provided. A step-by-step tutorial of developing solution strategies for energy systems optimization problem using both the above-mentioned quantum computing architectures is put forward.

- Novel reformulations of energy systems optimization problems namely facility location-allocation, unit commitment and heat exchanger network synthesis into unconstrained binary optimization problems has been proposed to facilitate ease of mapping and solving on quantum hardware.
- A performance comparison study in aspects of timing and solution quality is also conducted for each class of problems using both state-of-the-art quantum computing algorithms and best available classical CPU-based solution methods.
- While addressing the limitations of proposed methodologies owing to technological barriers, the future scope of quantum computing for large-scale complex energy systems optimization problems is discussed and some ideas to harness the full potential of quantum computers which can pave the way to a new sophisticated solution architecture for energy systems optimization are also put forth.

## 2. Quantum Computing Fundamentals

Over the past five decades, Moore's law has held true with the computing power doubling every two years by reducing the size of transistors in an integrated circuit. The current state of the art transistors are only a few atoms thick [27]. However, as this size gets smaller quantum effects start to interfere with their functionality. Quantum computing is the next frontier for computation. A broad definition of quantum computing is computing that follows the logic of quantum mechanics.

Analogous to *bit* being the fundamental unit of classical computing, quantum computing is built upon the *quantum bit* called *qubit.* A classical bit can be in either of state 0 or 1, while a qubit is in the superposition state of $|0\rangle$ and $|1\rangle$ together termed as computational basis states. Dirac notation "$|\ \rangle$" is the standard notation for states in quantum mechanics. Fig. 2 shows the *Bloch sphere* representation of a quantum bit which helps to visualize the state of a single qubit. Unlike classical bit which can be in either of its two states, infinite number of states are possible for a qubit. After it is measured, the state of qubit collapses to one of the basis states [11]. Another elegant property of qubits which sets them apart from classical bits is entanglement. Quantum entanglement is the ability to form co-relations between individually random behaviors of two qubits. These properties of superposition and ability to exist in entangled states are exploited in a quantum computer resulting in high computation power. A few of proposed quantum computing architectures have already been realized through technological advancements. A brief overview on the operation and the applications of these quantum computing technologies is provided below.

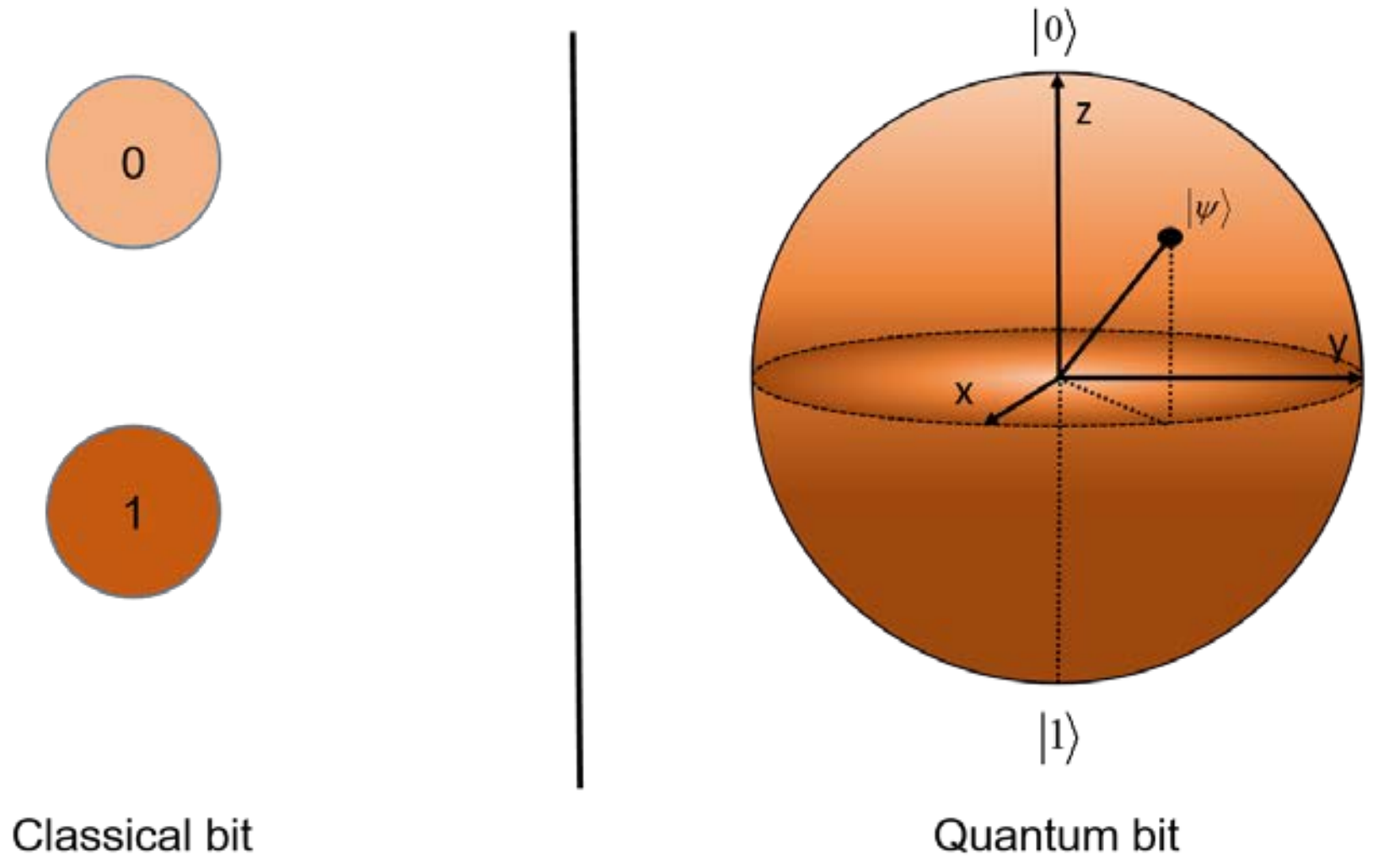

*Fig. 2. Fundamental units of computing*

## 2.1. Quantum Circuit Model

A quantum circuit comprises of quantum gates manipulating qubits to perform calculations. Analogous to logic gates in classical digital circuits, quantum gates are the building blocks of quantum circuits. In this model, quantum gate operations are applied one by one to the state of the system, thus evolving it towards a desired solution of the problem [28]. Gate model of quantum computation is an alternate term used for this model. In the quantum circuit, the tasks of preparing a specific input, applying a set of gate operations and measuring the state of qubits in the computational basis is carried out sequentially [29]. Quantum algorithms play a vital role in implementing the power of quantum hardware. Approximate solutions for combinatorial optimization problems can be produced by Quantum Approximate Optimization Algorithm (QAOA) [30]. In the field of chemistry and optimization, a classical-quantum hybrid algorithm called Variational Quantum Eigensolver (VQE) has shown exceptional performance in terms of speed and resource utilization [31]. Generating a trial state and estimating its energy is performed on a quantum computer with the energy being optimized on classical computer.

However, practical applications of the gate model quantum computer are limited. Few problems like integer factorization and solving linear system of equations show exponential speedup [32], while for other problems gate model quantum computers are not known to be faster than CPU/GPU-based classical computers. The qubits in this model are susceptible to de-coherence, meaning their quantum states are destroyed by interactions with the environment. A universal quantum computer can be realized after overcoming these setbacks and joining forces with classical computing. Tech giants in the field of computation like IBM have already launched a cloud based commercial version of a gate model quantum computer, and is working towards building an universal quantum computer [33].

## 2.2. Quantum Annealing based model

Quantum Annealing, which is essentially the noisy version of quantum adiabatic computation, is a quantum counterpart of simulated annealing. Instead of thermal fluctuations [7], quantum fluctuations are introduced in quantum annealing for faster convergence to optimal state [34]. A quantum annealing machine has wide range of applications, including ability to solve discrete/combinatorial optimization problems. Unlike the gate circuit model, quantum annealing machines or annealers are explicitly designed for a specific purpose, like optimization, machine learning, *etc*. It has also been proven that an extended form of quantum annealing is theoretically equivalent to gate (circuit) model [35]. The Canadian company, D-Wave, provides state of the art quantum annealing machines that have wide range of applications with cloud access to its users [36].

This model of quantum computer realizes the quantum annealing algorithm by enabling the quantum tunneling processes. In order to minimize the cost function, a time dependent term representing quantum fluctuations is added, such that the probability of existence of optimum is uniform over all possible states. This can be seen in Fig. 3 where the cost function is a double well with local optimum states and the probability distribution $P(x,t)$ being uniform at the beginning. After allowing the natural time evolution of the system, the probability is highest at global optimal solution at the end of the annealing process. Apart from quantum hardware, this algorithm can also be simulated on classical computers using a deterministic approach or quantum Monte Carlo techniques for large problems [37].

Although quantum annealing machines are able to withstand noise from environment, there are very few problems, which can be solved exponentially faster on these machines as compared to classical algorithms. Also, in comparison to gate circuit model the number of qubits used in annealers is much higher. Simultaneously, a hardware and language independent approach is being developed to enable migration of existing scientific code across all architectures [38]. At this stage of development there is a tradeoff between the above mentioned models of quantum computation and an informed decision must be made based on the complexity class, size of the problems, and whether or not a quantum advantage is expected.

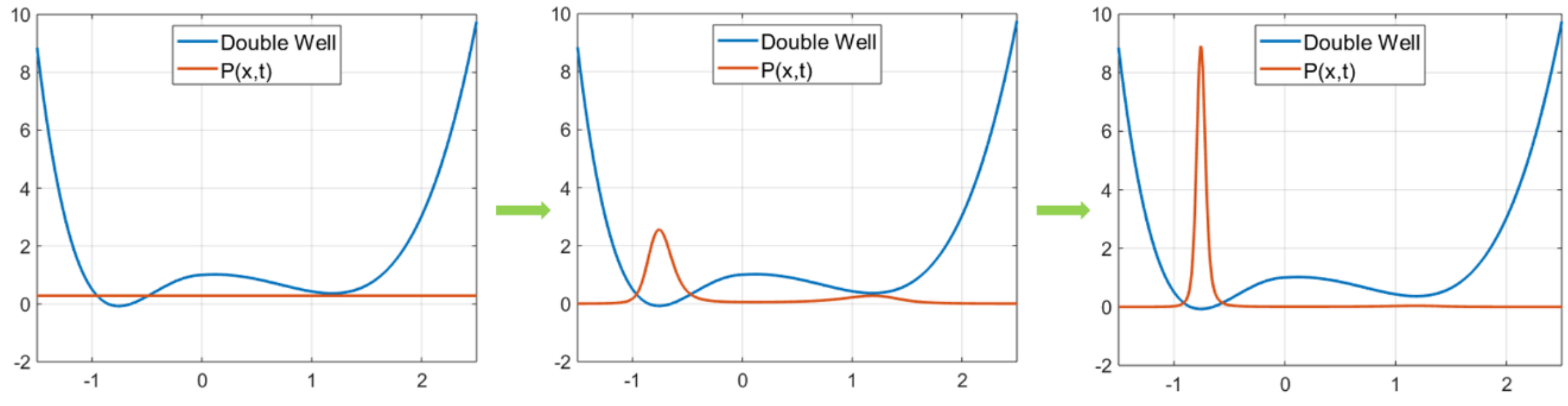

*Fig. 3. Time evolution of Quantum Annealing process for optimization of a function having both local and global optimum*

## 3. Facility location-allocation in energy systems optimization

Facility location and allocation problems are essential components of strategic design and planning for energy systems [39]. Some energy systems infrastructure development optimization problems could be cast as facility location-allocation problems. There has been substantial research on their model formulations, and various algorithms have been developed to solve such problems [40]. Examples of this type of energy systems optimization problems include determining the optimal location for wind farms to maximize energy capture while accounting for electrical grid constraints [41]. Locations of facilities such as solar or wind power generation systems or hydro power plant can also be determined based on minimization of facility opening costs, energy transportation costs subject to satisfying energy demand subject to resource availability constraints [39].

One such fundamental problem from the category of facility location is the quadratic assignment problem [42]. Hub-based network design for electricity storage require a hub facility location model, which is formulated as a quadratic assignment problem [43, 44]. Some solution approaches for biofuel supply chain optimization also use quadratic programming models [45]. The quadratic assignment problem being one of the most difficult problems of the NP-hard complexity class, is widely studied in the combinatorial optimization community. A number of formulations and solution techniques have been proposed accounting for the continuous interest in this problem [46]. In the context of power grid optimization, $n$ power plants are to be assigned to $n$ regions in such a way that the cost of interplant transportation is minimized. This is illustrated for four plants or facilities in Fig. 4. Two $n \times n$ matrices $C=||C_{ij}||$and $T=||T_{pq}||$ are given, where $C_{ij}$ is the cost of transporting one unit of energy from location $i$ to location $j$ and $T_{pq}$ is the number of units of energy to be transported from plant $p$ to plant $q$. Each of the $n^2$ assignment variable is denoted by binary variable $x_{pi}$ to represent whether plant $p$ is assigned to location $i$. The assignment constraints ensure that only one facility is assigned to each location. For simplicity purposes, any demand constraints and resource availability constraints are not considered here. The Koopmans-Beckmann formulation [47] of this problem is given in Eq. (1).

$$\begin{aligned} \min \quad & \sum_{q=1}^{n}\sum_{p=1}^{n}\sum_{j=1}^{n}\sum_{i=1}^{n} C_{ij} T_{pq} x_{pi} x_{qj} \\ \text{s.t.} \quad & \sum_{p=1}^{n} x_{pi} = 1, \ \forall i = 1, 2 \ldots n \\ & \sum_{i=1}^{n} x_{pi} = 1, \ \forall p = 1, 2 \ldots n \end{aligned} \tag{1}$$

In this section, the ability of different quantum hardware, namely the IBM's gate model and D-Wave's quantum annealers, to solve this combinatorial optimization problem is demonstrated. Problem instances for quadratic assignment were borrowed from the quadratic assignment problem library (QAPLIB), which provides a unified testbed for quadratic assignment problems through a compilation of their best known feasible solutions and best lower bounds [48]. These problem instances are solved on the D-Wave's 2000Q quantum processor, and its performance is compared to that of a deterministic solver Gurobi run on a Dell Optiplex system with Intel(R) Core(TM) i7-6700 3.40GHz CPU and 16 GB of installed memory. The same system configuration and computing environment was used to solve all of the problem instances mentioned henceforth for both quantum and classical computers.

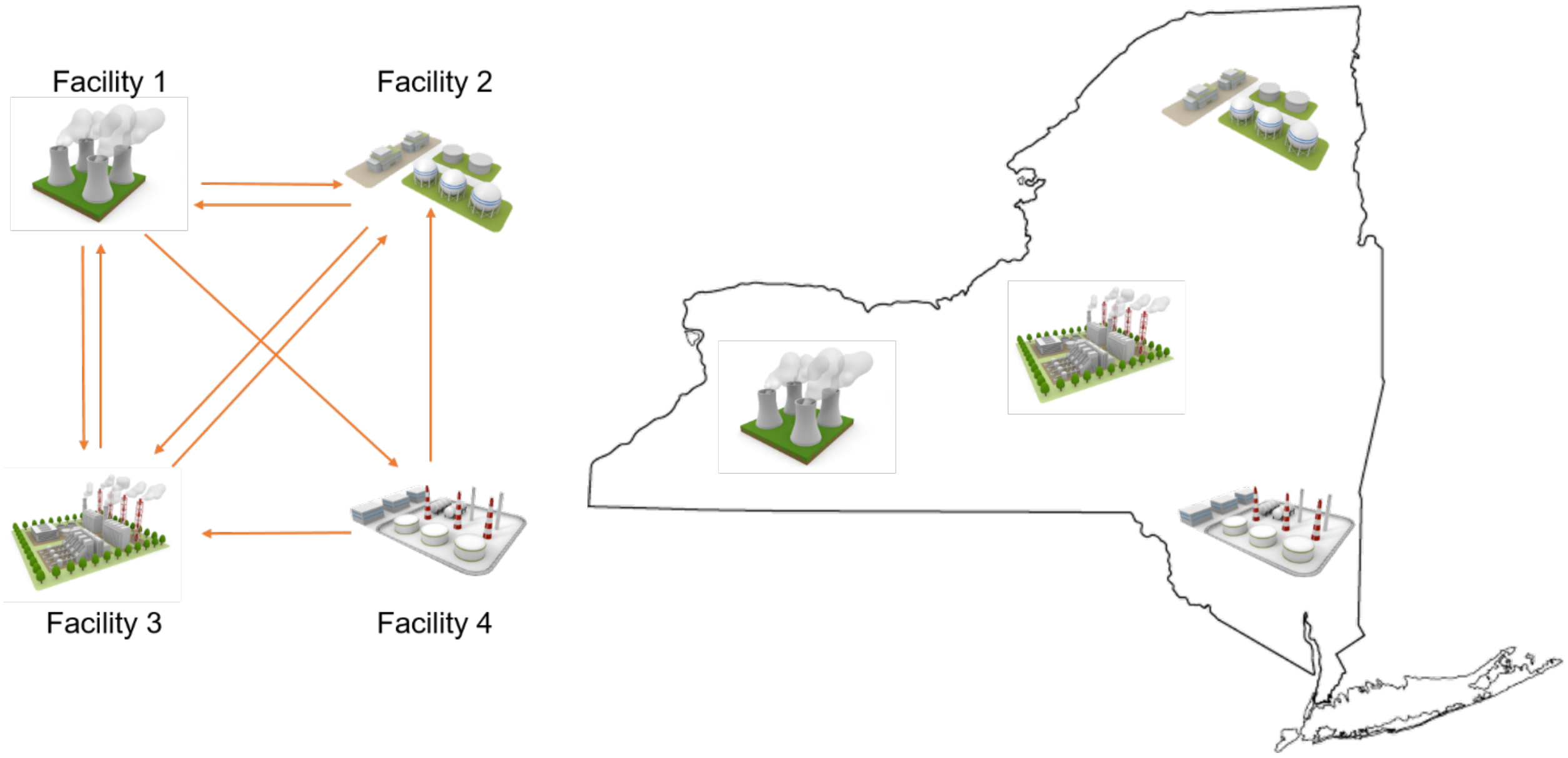

*Fig. 4. Flow between facilities and their locations*

## 3.1. IBM Q's quantum computer

IBM Q is an initiative to develop scalable quantum systems with a long term goal to realize an universal quantum computing system [49]. Their devices can be accessed through an open-source quantum computing framework called Qiskit [50]. Qiskit has three primary components, Terra, Aqua, and Aer, and they provide for composing quantum programs at circuit-level, tools and libraries for building quantum applications, and a high-performance simulator for quantum circuits, respectively.

The quadratic assignment problem can be converted into a single objective cost function given by Eq. (2) where the parameter $A$ is a free parameter chosen to be large enough to ensure constraint satisfaction. In order to map the problem to a quantum computer, an Ising Hamiltonian corresponding to the objective function must be minimized. The Ising Hamiltonian takes the form of Eq. (3) where $Z_i$ is a Pauli operator assuming the eigenvalues $\pm 1$, analogous to a binary variable taking values in $\{-1, 1\}$. The mapping of objective function to an Ising Hamiltonian is done by the assignment $x_{ij} \rightarrow (1 - Z_{ij})/2$, which yields the final form in Eq. (4).

$$C(x) = \sum_{q=1}^{n}\sum_{p=1}^{n}\sum_{j=1}^{n}\sum_{i=1}^{n} C_{ij} T_{pq} x_{pi} x_{qj} + A\sum_{i}\left(1 - \sum_{p} x_{pi}\right)^2 + A\sum_{p}\left(1 - \sum_{i} x_{pi}\right)^2 \quad (2)$$

$$\mathrm{H} = \sum_{i} w_i Z_i + \sum_{i<j} w_{ij} Z_i Z_j \tag{3}$$

$$\begin{aligned} C(Z) = \sum_{q=1}^{n}\sum_{p=1}^{n}\sum_{j=1}^{n}\sum_{i=1}^{n} C_{ij} T_{pq} \left(\frac{1-Z_{pi}}{2}\right)\left(\frac{1-Z_{qj}}{2}\right) + A\sum_{i}\left(\sum_{p}\left(\frac{1-Z_{pi}}{2}\right)-1\right)^2 \\ + A\sum_{p}\left(\sum_{i}\left(\frac{1-Z_{pi}}{2}\right)-1\right)^2 \end{aligned} \tag{4}$$

A code sample written in Python programming language is shown in Fig. 5.a where a quadratic assignment problem instance is solved using a quantum algorithm. In case of custom user defined problems, the Ising Hamiltonian needs to be manually generated. Exploring the open-source codebase for the various Ising translators should serve this purpose. The VQE quantum algorithm is chosen here among all those available in the Aqua library. A classical optimization technique and a variational method to calculate approximate wave functions along with an entangler map which specifies the entanglement of qubits are selected, depending on their specific properties and relevance to the ultimate goal. Finally, the quantum algorithm can be run on a Qasm simulator or on IBM Q devices remotely via cloud [51].

Execution of the quantum algorithm produces a probability distribution for possible lowest energy states or ground states for the Ising Hamiltonian, which correspond to the optimal solutions of the original cost function. The solution with highest probability tends to be the most likely global optimum. A binary string corresponding to the solution and its probability is returned after execution. Fig. 5.b represents a histogram for this probability distribution, which clearly highlights the most probable solutions for their respective problem instances.

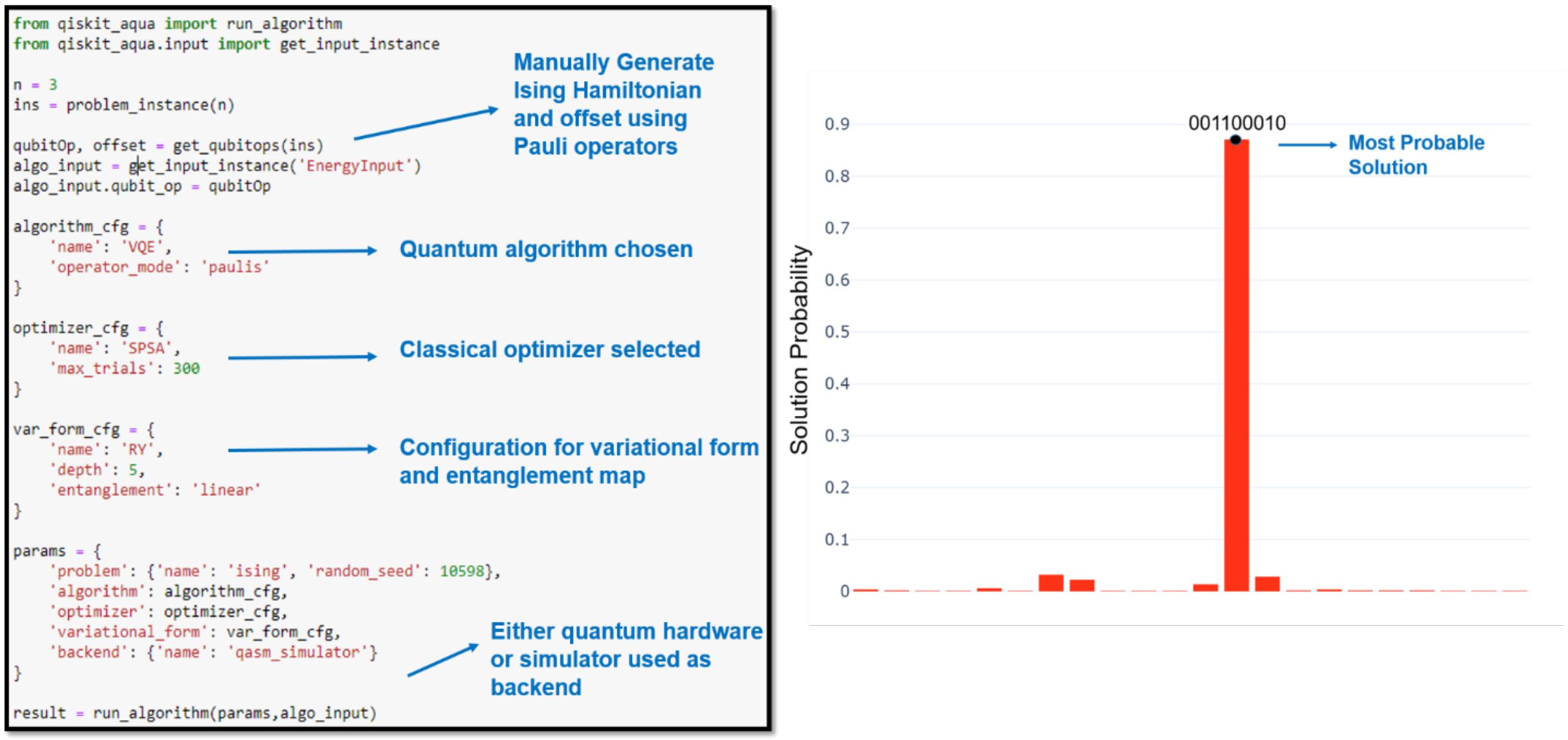

a) b)

*Fig. 5. a) Python code sample showing the use of Qiskit to solve a problem instance, and b) Histogram representing likelihood of possible solutions*

## 3.2. D-wave's quantum annealer

D-Wave systems use quantum annealing process to search for solutions to a problem. The D-Wave 2000Q system contains a quantum processing unit (QPU) with 2048 qubits. To solve hard problems with quantum computer, D-Wave provides a set of open-source Python tools called Ocean software development kit [52]. A problem along with the user specified parameters is submitted to the QPU, and problem solutions corresponding to the optimal configuration of qubits are returned over the network.

Only problems which can be mapped onto an Ising model in Eq. (4) or a quadratic unconstrained binary optimization (QUBO) formulation as in Eq. (2) can be solved on a D-Wave system. The coefficients of all the linear and quadratic terms corresponding to the QUBO form of a problem instance must be declared beforehand in Python dictionary format [53]. A sample code using the Ocean software libraries to solve the quadratic assignment problem instance and likelihood of best feasible solutions is shown in Fig. 6, where the 'linear' and 'quadratic' terms are user defined. Dimod is an utility that generates the binary quadratic model using both linear and quadratic terms provided [54]. Embedding of the problem by mapping it to physical qubits on the annealer is an important step here. Qbsolv is a decomposing solver which finds the minimum of a large QUBO by partitioning it into smaller sub-problems [55]. This solver can be configured to solve the QUBO problem instance by a meta-heuristic algorithm, Tabu Search, on a CPU-based classical computer or by quantum algorithm on D-Wave's quantum system.

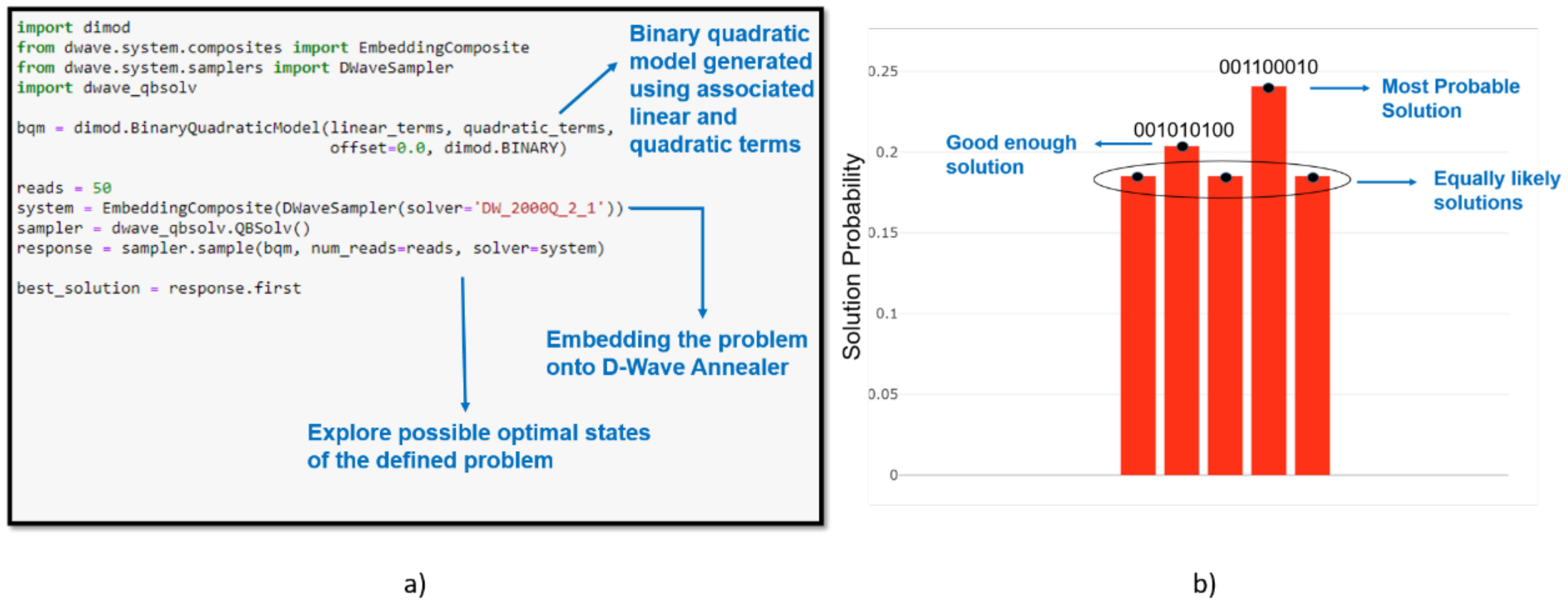

*Fig. 6. a) Python code sample showing use of Ocean software libraries to solve a problem instance, and b) Histogram representing likelihood of possible feasible solutions*

Since either of the Ising model or quadratic unconstrained binary optimization model is compatible with the D-Wave systems, the problem in Eq. (1) needs to be mapped appropriately. Further simplification of the single objective cost function in Eq. (2) and reducing $x^2$ to $x$, as $x$ being a binary variable belonging to $\{0,1\}$ yields the final QUBO in Eq. (5) with the value of parameter $A$ chosen, such that $A \gg \max\left(C_{ij}T_{pq}\right)$, and the constant term $2nA$ used as an offset. The QUBO is then mapped onto the Chimera graph architecture of qubits on D-Wave QPU through the minor-embedding process. The QPU then minimizes the energy of this configuration, thus locating the lowest energy state of this configuration which corresponds to the minimum of original cost function. The D-Wave annealers are explicitly built for optimization and sampling problems, so optimization problems tend to run faster on them than on classical computers. However, as problem size increases partitioning it into sub-problems might affect the timing performance and the quality of produced solution.

$$C(x) = 2nA + \sum_i \sum_j \sum_p \sum_q C_{ij} T_{pq} x_{pi} x_{qj} + 2A \sum_i \sum_p \sum_{q>p} x_{pi} x_{qi} + 2A \sum_p \sum_i \sum_{j>i} x_{pi} x_{pj} - 2A \sum_p \sum_i x_{pi} \qquad (5)$$

*Table 1. Model statistics of quadratic assignment instance for n facilities and n locations*

| | **KB formulation (Eq. 1)** | **QUBO (Eq. 5)** |
|---|---|---|
| no. discrete variables | $n^2$ | $n^2$ |
| no. constraints | $2n$ | 0 |

## 3.3. Results

Problem instances of size ranging from 3 to 20 facilities and locations were solved using both D-wave's quantum computer and CPU-based classical computer. Known optimal solutions are available on the QAPLIB collection [48] for the used problem instances. Solutions given by an exact deterministic solver Gurobi implemented on Intel(R) Core(TM) i7-6700 3.40GHz CPU were used for cross validation and for the purposes of benchmarking. Model sizes for quadratic assignment instances for $n$ locations and $n$ facilities are given in Table 1. This means that the largest problem with 20 candidate locations and 20 facilities includes 400 binary variables.

Table 2 shows the detailed computational times and objective function values for the quadratic assignment problem instances. All the instances were solved using both classical Gurobi solver and quantum Qbsolv solver, and the objective values along with computational times required to calculate them were noted. Best solutions obtained using the quantum solver are reported in the computational results. A timeout of 12 hours was set for the Gurobi solver, meaning solver algorithm would stop the solution process after the specified time limit. Objective function values at this time limit are reported. Best known solutions for these problem instances are also given in Table 2 to compare the quality of obtained solutions with both classical and quantum techniques.

As evident from Table 2, the classical solver Gurobi uses up exponentially more time as problem size increases. However, that is not the case with quantum solver. Smaller sized instances of the quadratic assignment problem which can be embedded onto the specific Chimera graph of the quantum processor are solved within less than 0.07 seconds. Time required by these small scale instances is more or less similar. This is because after the process of embedding the whole QUBO corresponding to a small problem instance on to the Chimera graph is finished, the computation time includes only the annealing and post-processing times which are proportional to the rate of reading samples. On the other hand, when an embedding cannot be found for a larger QUBO instance, the solver Qbsolv splits the input QUBO instance into sub-QUBOs that fit on the D-Wave system in use. This classical-quantum hybrid algorithm exploits the complementary strengths of both tabu search and D-Wave solver [56].

*Table 2. Computational results for solving quadratic assignment problem instances with classical and quantum solvers*

| No. facilities | Best known solution | Gurobi solver (single CPU core) | | Quantum solver (D-wave 2000Q) | |
|---|---|---|---|---|---|
| | | **time(s)** | **obj. fun.** | **time (s)** | **obj. fun.** |
| 3 | 24 | 1.33 | 24 | 0.024 | 24 |
| 4 | 32 | 1.48 | 32 | 0.062 | 32 |
| 5 | 58 | 1.5 | 58 | 0.066 | 58 |
| 6 | 94 | 1.35 | 94 | 0.043 | 94 |
| 8 | 214 | 1.96 | 214 | 0.127 | 214 |
| 9 | 264 | 2.01 | 264 | 445.23 | 264 |
| 12 | 578 | 325.68 | 578 | 1946.12 | 578 |
| 14 | 1014 | 42,010.42 | 1014 | 1008.7 | 1026 |
| 15 | 1150 | ---* | 1160 | 986.19 | 1160 |
| 17 | 1732 | ---* | 1750 | 921.71 | 1786 |
| 20 | 2570 | ---* | 2674 | 744.76 | 2640 |

* Timeout of 12 hours reached by Gurobi

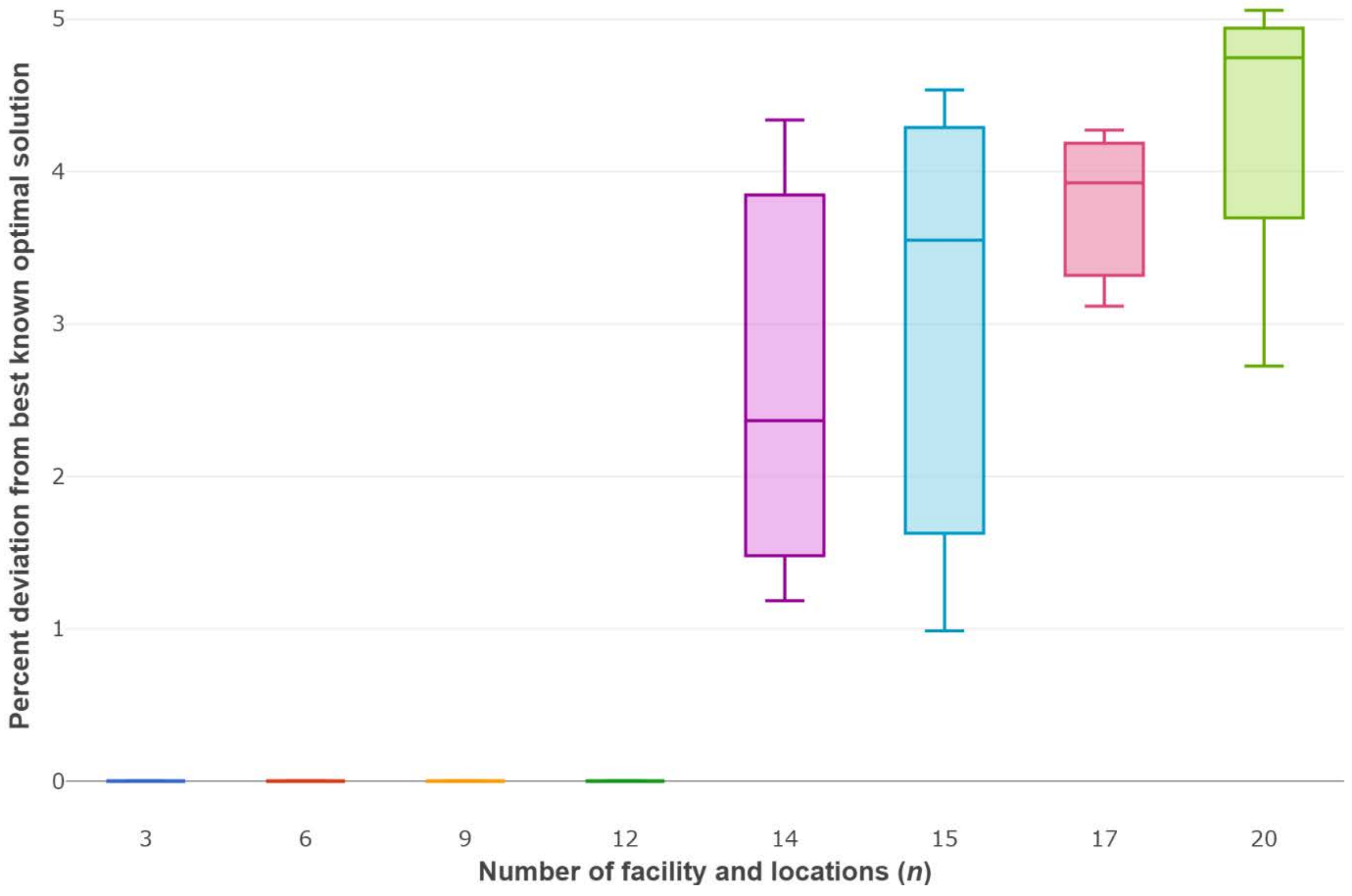

*Fig. 7. Solution quality for quadratic assignment problem instances obtained using quantum solver*

Similar to other heuristic techniques, a decline in quality of solution is observed for large problems, where solution space grows exponentially with the size. The quality of obtained solution can be represented as deviation from the best available optimal solution. This can be seen in Fig. 7, in which a distribution of percentage deviation from best known optimal solution for each problem instance is plotted. These results show that good solutions without excessive computation times can be obtained with quantum solver implemented on D-wave system. In spite of the percentage deviation being not greater than five percent, a more robust partitioning scheme might be able to eliminate or lower this deviation.

## 4. Unit commitment problem

Unit commitment (UC) is one of the most popular and critical optimization problems in the electrical power industry for the operations of power systems. The UC problem is generally formulated as a large-scale mixed integer nonlinear problem and solving it is very difficult due to the nonlinear cost function and the combinatorial nature of set of feasible solutions. It has been proven that UC is not only NP-hard but also NP-complete, so it is impossible to develop an algorithm with polynomial computation time to solve it [57]. Various techniques to solve the UC problem have been proposed based on deterministic approach [58], meta-heuristic methods and combinatorial approaches [59]. Efficient optimization techniques to solve advanced UC models when high renewable energy sources are integrated to the power systems, are of critical importance [60]. With the increasing number of energy sources, UC problem poses a hard challenge making it crucial to develop effective methodologies to tackle this challenge.

Unit commitment is concerned with minimizing the total operational cost to meet an estimated electric power demand over a given time horizon, while a number of system and generator constraints must be met [61]. Fig. 8 clearly illustrates the matching of combination of units to different loads or power demands in a given time interval. System power balance, spinning reserves and generation power limits of each unit account for some of the constraints in unit commitment. From a set of units $U$ the amount of power generated by each unit $i$ to satisfy a given load requirement $L$ is represented by $p_i$. The fuel cost $f_i$ of the committed unit $i$ is usually formulated as a quadratic polynomial with $A_i$, $B_i$, and $C_i$ being the coefficients of this polynomial. The UC problem for a single time period is mathematically represented by mixed-integer quadratic programming problem in Eq.(6), where binary variable $y_i$ represents whether the corresponding unit $i$ is online. The generated power limits of each unit $i$ are imposed by lower bound $P_{min,i}$ and an upper bound $P_{max,i}$. Here, randomly generated UC problem instances were used for study. After reformulation, these instances were solved on both D-wave 2000Q quantum processor and on Intel(R) Core(TM) i7-6700 3.40GHz CPU using Gurobi solver to compare the quality of obtained solutions.

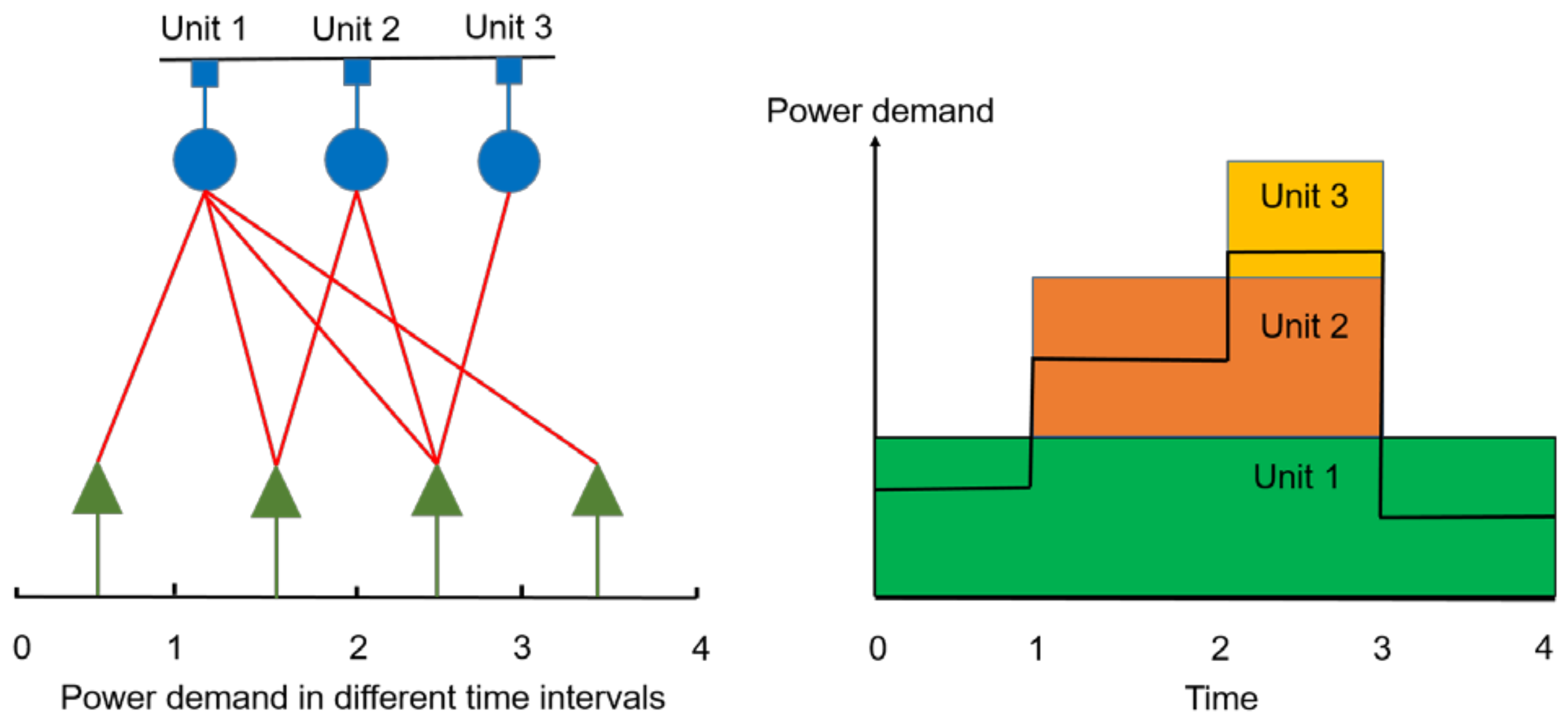

*Fig. 8. Commitment of units to different loads/power demands in each time interval*

$$\begin{aligned}
&\min \sum_{i\in U} f_i \\
&s.t.\quad f_i = A_i y_i + B_i p_i + C_i p_i^2 \\
&\qquad \sum_{i\in U} p_i = L \\
&\qquad P_{\min,i} y_i \le p_i \le P_{\max,i} y_i \qquad \forall i \in U \\
&\qquad y_i \in \{0,1\}
\end{aligned} \tag{6}$$

## 4.1. Implementation

Only an unconstrained quadratic optimization problem can be mapped on to D-wave system, so the mixed-integer quadratic programming problem in Eq. (6) cannot be directly solved on quantum computing system. The UC problem must be reformulated into a QUBO model, which can be achieved by discretizing the problem space. The feasible range of continuous variable $p_i$ is divided into $N$ equally spaced grids between $P_{min,i}$ and $P_{max,i}$ while ensuring that this variable equals zero if the unit is offline. Since this reformulation is an approximation of the original problem, the number of grids was chosen such that optimal solution of the reformulated UC is within 0.01% of exact global optimum obtained using a nonlinear solver. Since the problem in Eq. (6) is convex in nature, Gurobi solver is used to solve the mixed integer quadratic programming problem. The reformulated UC given in Eq. (7) is a quadratic binary model and can be easily converted to QUBO through a single objective cost function corresponding to the discretized problem. Eq. (8) forms the required QUBO and can hence be embedded on to the D-wave quantum processor. The same instance is solved multiple times on D-wave QPU to compare the likelihood of obtained solutions to be a global optimum and choose the best solution.

$$
\begin{aligned}
\min \ & \sum_{i \in U} f_i \\
s.t. \quad & h_i = \left( \frac{P_{\max,i} - P_{\min,i}}{N} \right) \\
& p_i = \sum_{k=1}^{N+1} \left( P_{\min,i} + (k-1) h_i \right) z_{ik} \\
& f_i = A_i (1 - v_i) + B_i p_i + C_i p_i^2 \\
& v_i + \sum_{k=1}^{N+1} z_{ik} = 1 \qquad \forall i \in U \\
& \sum_{i \in U} p_i = L \\
& v_i, z_{ik} \in \{0,1\}
\end{aligned}
\tag{7}
$$

$$
\begin{aligned}
C(y,z) = & \sum_{i \in U} \sum_{k=1}^{N+1} \left( A_i + B_i \left( P_{\min,i} + (k-1) h_i \right) + C_i \left( P_{\min,i} + (k-1) h_i \right)^2 \right) z_{ik} \\
& + \sum_{i \in U} \sum_{k=1}^{N+1} 2 \mathrm{C}_i \left( P_{\min,i} + (k-1) h_i \right) \left( P_{\min,i} + (m-1) h_i \right) z_{ik} z_{im} \\
& + A \sum_{i \in U} \left( v_i + \sum_{k=1}^{N+1} z_{ik} - 1 \right)^2 + B \left( L - \sum_{i \in U} \sum_{k=1}^{N+1} \left( P_{\min,i} + (k-1) h_i \right) z_{ik} \right)^2
\end{aligned}
\tag{8}
$$

## 4.2. Results

The model sizes for unit commitment problem with $n$ units and $N$ grids are given in Table 3. The size of discretized problem increases with the number of grids $N$. For the largest problem with 12 units, the original nonlinear problem includes 12 binary variables, 12 continuous variables and 25 constraints, while the reformulated problem includes 144 binary variables and only 13 constraints.

The randomly generated problem instances for unit commitment in a single time interval were solved using both classical and quantum solvers. The computational results for the same are given in Table 4, where the objective function value and time required are mentioned for each model and respective solution algorithm used. Best solutions and computational time obtained by quantum solver are also reported in this table. As mentioned earlier choosing proper value of $N$ ensured a good optimum solution lying within 0.01% of the global optimum. The number of grids used for discretization is also noted here for each problem instance. Optimal solution of nonlinear and discretized UC in Eq. (6) and (7), was calculated by the Gurobi solver using regular CPU on Intel(R) Core(TM) i7-6700 3.40GHz CPU. For the quantum solver, the QUBO parameter values $A$ and $B$ in Eq. (8) vary for each instance and are chosen based on the coupling strengths and biases in the original cost objective function and usually take high values to ensure constraint satisfaction. Biases and coupling strengths are simply the linear and quadratic terms in the QUBO. Values of $A$ and $B$ are also reported for each instance. The smallest discretized instance with 27 binary variables can be embedded on to Chimera architecture of quantum processor with ease. However, as the size of problem and number of grids increase, the corresponding QUBO cannot be directly embedded on the D-wave QPU. Larger QUBOs require partitioning into smaller sub-problems or sub-QUBOs, and can be solved using the Qbsolv utility. The partitioning scheme reduces the likelihood of returning global optimal solutions and might return good enough solutions. Hence, for large instances, solutions obtained through quantum algorithm exhibit deviation from solutions obtained using the global optimization solver Gurobi.

*Table 3. Model statistics for unit commitment problem with n units and N grids*

| | **Non-linear model Eq. (6)** | **Discretized model Eq. (7)** | **QUBO Eq. (8)** |
|---|---|---|---|
| Number of discrete variables | $n$ | $n(N+2)$ | $n(N+2)$ |
| Number of continuous variables | $n$ | $0$ | $0$ |
| Number of constraints | $2n+1$ | $n+1$ | $0$ |

*Table 4. Computational results for solving unit commitment problem using both classical and quantum solvers*

| | Gurobi solver (single CPU core) | | | | | | | | Quantum solver (D-wave 2000Q) | |
|---|---|---|---|---|---|---|---|---|---|---|
| | Non-linear model | | | | Discretized model | | | | | |
| **no. units** | **obj. fun.** | **time (s)** | **Grids** | **% Error** | **obj. fun.** | **time (s)** | **A** | **B** | **obj. fun.** | **time (s)** |
| 3 | 25.5 | 0.024 | 7 | 0.00% | 25.5 | 0.016 | 2000 | 5 | 25.5 | 0.03 |
| 5 | 13.9 | 0.028 | 10 | 0.00% | 13.9 | 0.02 | 1000 | 10 | 13.9 | 0.07 |
| 8 | 22.5 | 0.026 | 10 | 0.00% | 22.5 | 0.022 | 3000 | 5 | 25.8 | 385.75 |
| 12 | 44.49 | 0.036 | 10 | 0.01% | 44.5 | 0.03 | 6000 | 5 | 53.35 | 1138.16 |

For a large discretized UC problem solved with Gurobi, increasing the number of grids reduces the deviation from global optimal solution. Ideally, a similar behavior should be observed for the case of quantum solver. But high number of grids require higher precision on the quantum system. Due to the precision limitations for biases and coupling strengths on the D-wave quantum system, discretizing problem space into smaller grid sizes does not guarantee a better solution. Dividing the space into grids beyond a certain limit would be redundant and solution quality would further deteriorate. Thus, there is a trade-off between precision and likelihood of global optimal solution when solving quantum optimization problems.

## 5. Heat exchanger network synthesis

In order to avoid excessive energy consumption in industrial plants, energetic integration of processes is necessary. Heat exchanger network synthesis (HENS) is a systematic tool to control the costs of energy for a process. Heating and cooling costs are controlled by hot and cold utilities as well as the heat exchangers in a network. Enhancing heat transfer in heat exchanger devices using various techniques can also contribute to reduction of these costs and ensure efficient energy conversion [62, 63]. A cold stream in a process can be either heated by hot utility or hot stream in the process while a hot stream can be cooled by cold utility or another cold stream. Heat exchanger network in Fig. 9 comprising of 2 hot streams and 3 cold streams integrated with 4 heat exchangers illustrates this. Owing to the need for improvement in energy recovery and minimizing the global cost, heat exchanger network synthesis problem has been extensively studied in the past [64]. HENS is a NP-hard problem for which an exact polynomial-time computational algorithm does not exist [65]. Several metaheuristic techniques capable of handling complex problems in a feasible computational time have also been proposed [66, 67].

Sequential synthesis and simultaneous synthesis are the two well-known groups of methods for handling HENS problem. One of the prevalent approaches is sequential synthesis, wherein a problem is decomposed into three subproblems, namely the minimum utility cost, minimum number of matches, and minimum cost network problems. These subproblems can be treated in an easier fashion as compared to the original single-task problem. The minimum number of matches problem consists of determining minimum cost of matches, which satisfy the supply and demand of heat for a heat exchanger network. This simple HENS subproblem with only one temperature interval is NP-hard in the strong sense [65]. For a heat exchanger network consisting of $m$ sources and $n$ sinks with a single temperature interval, there is a supply $S_i$ of heat at each source $i$ and demand $D_j$ for heat at each sink $j$. The cost for each possible match between a source $i$ and a sink $j$ is denoted by $c_{ij}$. The variables $q_{ij}$ represent total heat exchanged between source $i$ and sink $j$, while the existence of heat match between them is given by a binary variable $w_{ij}$. Fig. 9 clearly shows the existence of heat exchanger match and heat flow between hot stream $i$ and cold stream $j$. This matches problem can be formulated as in Eq. (9) where the first two constraints represent energy balance, and the third one is a logical constraint. Here a randomly generated problem instance of size $m$=4 and $n$=3 was used for analysis that satisfied the feasibility requirement $\sum_i S_i = \sum_j D_j$. A pictorial depiction of the same is shown in Fig. 10a. This problem is reformulated and solved on D-wave's 2000Q quantum processor and on Intel(R) Core(TM) i7-6700 3.40GHz CPU using Gurobi solver for solution quality comparison.

$$
\begin{aligned}
\min \ & \sum_i \sum_j c_{ij} w_{ij} \\
s.t. \ & \sum_j q_{ij} = S_i \quad \forall i = 1, 2 \ldots m \\
& \sum_i q_{ij} = D_j \quad \forall j = 1, 2 \ldots n \\
& U_{ij} = \min\{S_i, D_j\} \\
& q_{ij} \le U_{ij} w_{ij} \quad \forall i, j \\
& q_{ij} \ge 0 \quad \forall i, j \\
& w_{ij} \in \{0,1\} \quad \forall i, j
\end{aligned} \tag{9}
$$

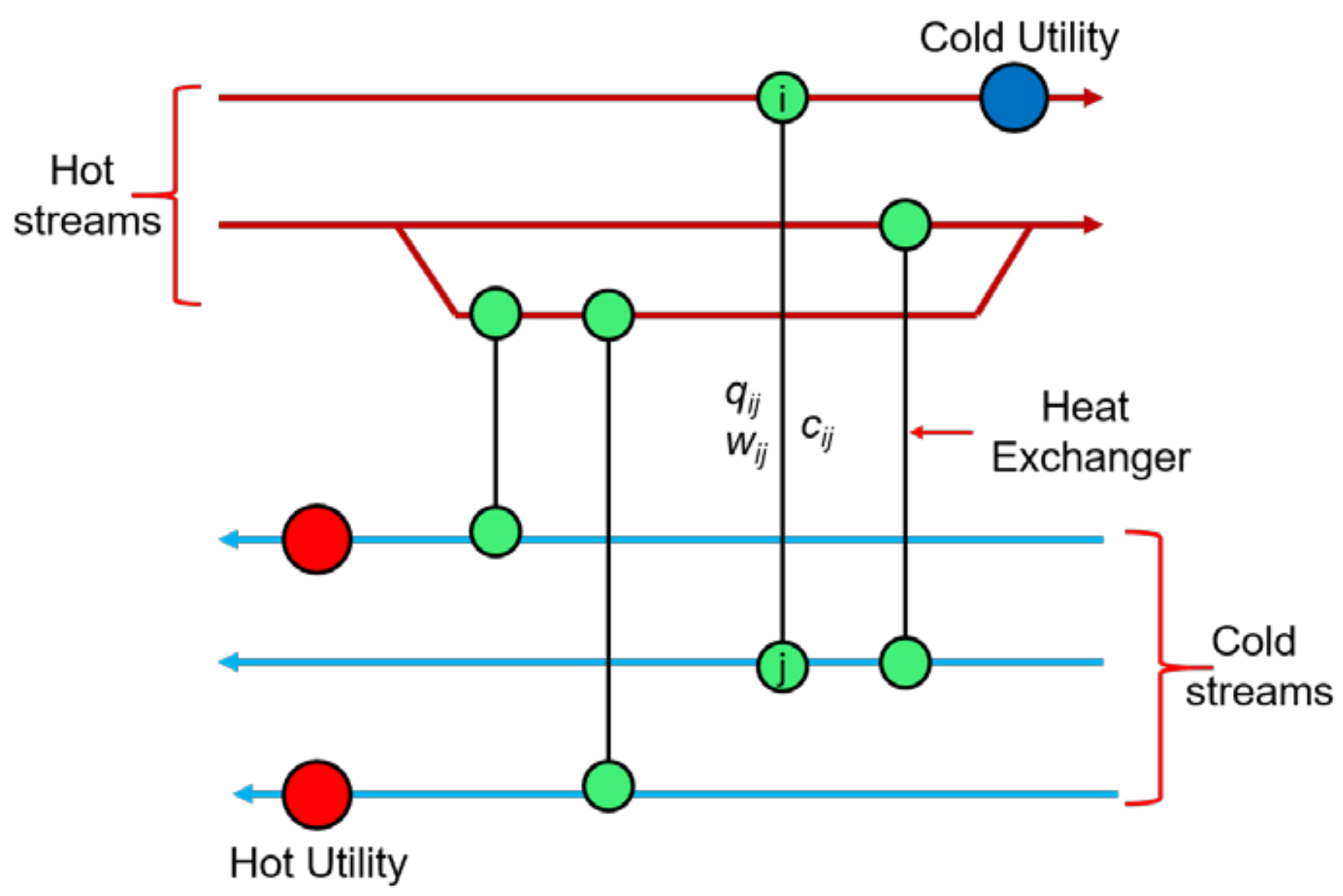

*Fig. 9. Heat exchanger network containing 2 hot streams and 3 cold streams*

## 5.1. Implementation

The minimum number of matches problem cannot be directly mapped on a quantum computer. As quantum computers support only discrete problems, the continuous variables $q_{ij}$ are discretized into $N$ equally-spaced grid points in the interval $[0, U_{ij}]$. The upper bound on $q_{ij}$ is set such that the logical constraint is not violated. Both the original problem and this discretized problem given by Eq. (10) is solved using the same Gurobi solver to check the viability of the reformulation. The discrete optimization problem can now be mapped on to D-wave QPU by converting it into a quadratic unconstrained binary optimization problem. For QUBO a single objective cost function is given by Eq. (11) containing only binary variables $z$. Here the value of $N$ was chosen such that both the mixed-integer linear problem (MILP) and discretized formulations in Eq. (9) and Eq. (10) respectively produce the optimal objective values with least deviation between them. QUBO corresponding to the same discretized problem instance is solved on D-wave QPU repeatedly and compared to the exact solution produced by the classical deterministic optimization solver Gurobi on a CPU-based computer.

$$
\begin{aligned}
\min\ & \sum_i \sum_j c_{ij} w_{ij} \\
s.t.\ & \sum_j q_{ij} = S_i \quad \forall i = 1,2...m \\
& \sum_i q_{ij} = D_j \quad \forall j = 1,2...n \\
& U_{ij} = \min\{S_i, D_j\} \\
& q_{ij} = \frac{U_{ij}}{N} \sum_{k=1}^{N} k z_{ijk} \quad \forall i, j \\
& \sum_{k=1}^{N} z_{ijk} = w_{ij} \quad \forall i, j \\
& z_{ijk}, w_{ij} \in \{0,1\} \quad \forall i, j, k
\end{aligned} \tag{10}
$$

$$
\begin{aligned}
C(z) = & \sum_i \sum_j c_{ij} w_{ij} + A\left( w_{ij} - \sum_{k=1}^{N} z_{ijk} \right)^2 + B \sum_i \left( S_i - \sum_j \sum_{k=1}^{N} \frac{U_{ij}}{N} k z_{ijk} \right)^2 \\
& + B \sum_j \left( D_j - \sum_i \sum_{k=1}^{N} \frac{U_{ij}}{N} k z_{ijk} \right)^2
\end{aligned} \tag{11}
$$

## 5.2. Results

Table 5 shows the statistics for model sizes of given formulations of minimum number of matches problem. For the randomly generated problem instances, MILP model and the discretized model are solved using CPU-based Gurobi solver. Its corresponding QUBO is solved on quantum solver and the computation times along with obtained solution are reported in Table 6. For the largest instance with 15 sources and 15 sinks, the original MILP problem consists of 225 continuous and binary variables while the discretized reformulation has 1350 binary variables. The number of grid points *N* was chosen such that the discretization scheme produced an optimal solution close to the one obtained by solving the original minimum number of matches problem using an exact solver Gurobi. Choosing the value of *N* also depends on the viability of solving very large QUBOs on quantum hardware. Fixing the number of grid points N ensures the existence of optimal solution close to global minima in the discretized problem space. Since the large instances cannot be directly embedded on to the Chimera graph architecture of quantum processor, the decomposing solver Qbsolv is used to partition the large QUBO into smaller sub-QUBOs. Values of parameters *A* and *B* of QUBO in Eq. (11) are set by trial-and-error, while satisfying $A, B \gg \max(c_{ij})$. Here *A*=20 and *B*=5 are used.

As evident from the computational results, smaller problem instances solved on quantum hardware return global optimal solutions. The best obtained solution shown in Fig. 10b exactly matches the one produced by Gurobi for discretized UC on Intel(R) Core(TM) i7-6700 3.40GHz CPU. For the largest problem instance, larger values of penalty weight B in the QUBO tend to violate the energy balance constraints. Larger HENS problem instances grow quickly with the number of grids and tend to perform poorly given the limited scalability of available quantum hardware. This is due to the fact that a finite precision is available on quantum annealers, and large $B$ values result in large coupling strengths and biases. Scaling down these strengths becomes necessary resulting in loss of precision, thus leading to constraint violation. A simple workaround for this issue would be to use appropriate number of grids. Also, sharing the edge weight of Chimera graph can enhance the precision at the expense of extra qubits. A small scale MILP model translates to a large discretized model, so Gurobi solver performs better than quantum solver with respect to time and solution quality. Partitioning of the problem into sub-QUBOs also reduces the likelihood of obtaining global optimal solutions. Improvement in solving large-scale problems on quantum processor is highly dependent on the scalability and precision offered by quantum systems. Alternative embedding schemes for QUBO on to quantum processor may facilitate some form of error mitigation thus improving solution quality. MILPs on CPU-based Gurobi solver have been optimized to run through parallel computation and perform exceptionally on conventional computers. Quantum computers on the other hand due to their small scalability and limited precision do not perform as good as classical MILP solvers. In the context of MILP solution approaches, quantum computers will require much more technological maturity to compete against their highly evolved classical counterparts.

*Table 5. Model statistics for minimum matches problem with m sources, n sinks and N grid points*

| | **MILP model Eq. (9)** | **Discretized model Eq. (10)** | **QUBO Eq. (11)** |
|---|---|---|---|
| Number of discrete variables | $mn$ | $mn(N+1)$ | $mn(N+1)$ |
| Number of continuous variables | $mn$ | $0$ | $0$ |
| Number of constraints | $m+n+2mn$ | $m+n+mn$ | $0$ |

*Table 6. Computational results for solving heat exchanger network synthesis problems using both classical and quantum solvers*

| | | Gurobi solver (single CPU core) | | | | | | Quantum solver (D-wave 2000Q) | |
|---|---|---|---|---|---|---|---|---|---|
| | | MILP model | | | | Discretized model | | | |
| no. sources | no. sinks | obj. fun. | time (s) | Grids | Error* | obj. fun. | time (s) | obj. fun. | time (s) |
| 2 | 2 | 5 | 0.005 | 4 | 0.00% | 5 | 0.009 | 5 | 0.03 |
| 3 | 3 | 9 | 0.015 | 4 | 0.00% | 9 | 0.009 | 9 | 0.08 |
| 4 | 3 | 9 | 0.009 | 5 | 0.00% | 9 | 0.014 | 9 | 261.6 |
| 10 | 9 | 20 | 0.025 | 5 | 0.00% | 20 | 0.058 | 21 | 519.2 |
| 15 | 15 | 26 | 0.049 | 5 | 0.00% | 26 | 0.106 | 26 | 1288.3 |

*Relative error between solutions obtained by discretized model and the global optimum

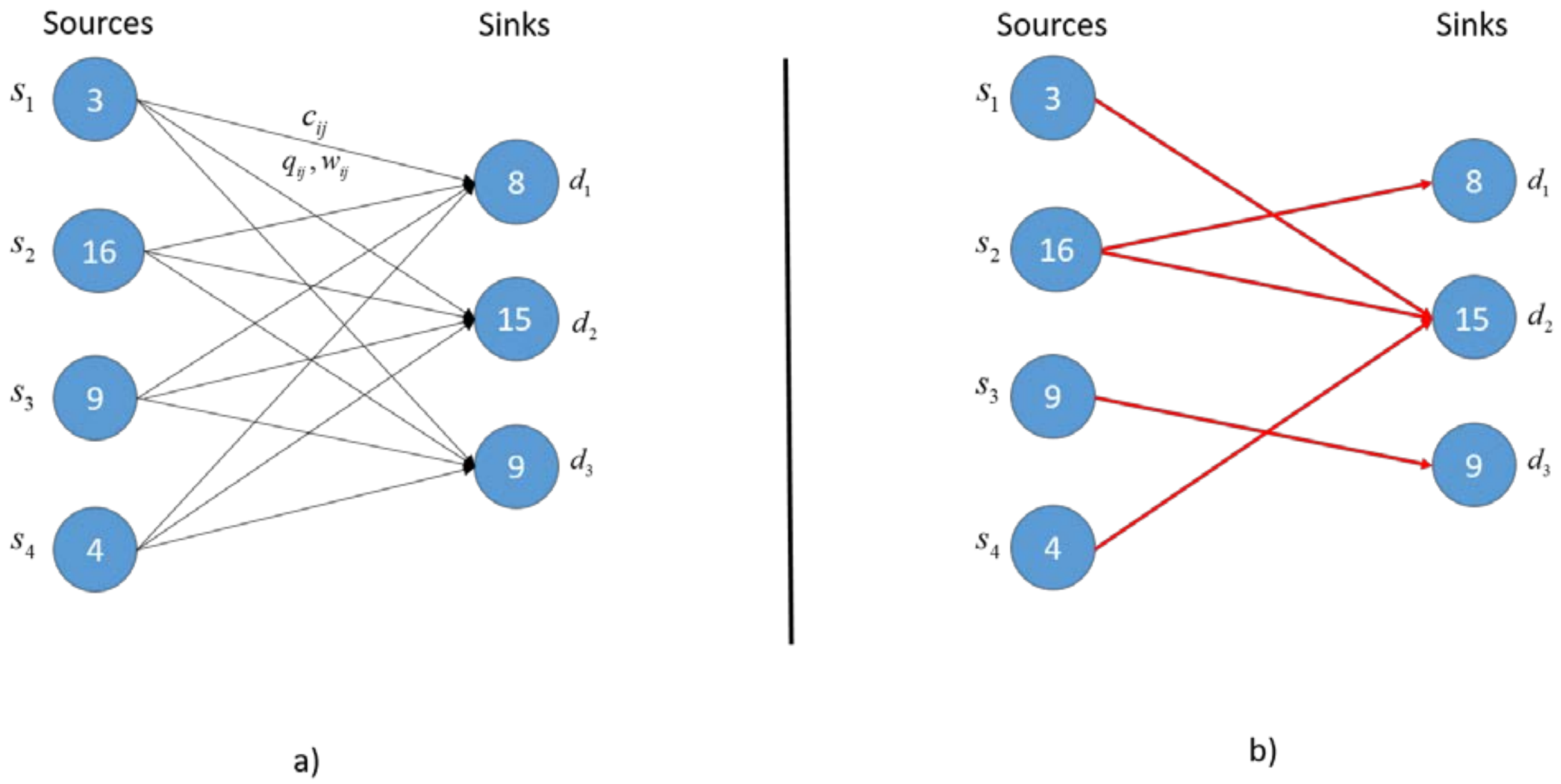

*Fig. 10. a) Minimum number of matches problem and b) Obtained optimal solution for the problem instance*

## 6. Discussion

D-wave quantum devices exhibit limitations in terms of problem size, connectivity of qubits and precision. The full speed of the machine can only be implemented, if the decision variables in a problem can be mapped to nodes on Chimera graph architecture of QPU. This process called minor embedding limits the scope of D-wave quantum machine. Minor embedding in itself maybe a NP-complete problem and would require an exponential preparation steps even if the hardware solved a problem in polynomial time. The limited connectivity offered by the QPU demands decomposing a large problem into smaller ones without a guarantee of best solutions. Unlike classical CPU/GPU-based computers, D-wave devices have very limited precision and this could cause scaling issues. Presence of intrinsic control errors in these systems and their lack of error correction negatively impacts the performance of quantum annealers. While quantum annealing based devices have shown performance gains for specific applications, algorithms for classical computers can be usually optimized for a specific problem enabling them to outperform quantum annealers at the current stage. This is evident from the large quadratic assignment instance performing better than a simpler heat exchanger network synthesis problem in terms of speed and solution quality. Even in case of a problem for which quantum speedup is apparent, there is no denying the possibility that a better classical algorithm will be found for the same class of problems.

The gate based quantum systems achieved by firms like IBM have very few qubits compared to the annealing systems. Different error sources add noise to qubits in turn affecting the robustness of gate operation. Current state of the art gate based quantum computer contains 20 qubits with gate error rate of about five percent. It is projected that for a quantum computer to achieve some task that no classical computer can perform will need over fifty qubits with less than 0.1 percent gate error rate. Improving the quality of qubits and qubit operations, as the number of qubits and associated controls rise, is a major challenge. With increasing number of qubits and their variance, an important goal would be to mitigate noise from different sources. It is expected that this challenge would likely get harder as the system size increases. At the current state of technological advancement, classical algorithms dominate the quantum algorithms implemented on gate systems with very few exceptions. On the other hand, quantum annealers developed explicitly for optimization perform better in some cases. This does not necessarily imply that simpler problems easily solvable on classical computer will perform exceptionally better on quantum hardware. Problems relevant to practical applications do not always belong to the class of binary quadratic model, rather a huge chunk of them are mixed-integer nonlinear optimization problems. Until a universal quantum computer is realized, it is only logical to use classical and quantum hardware in conjunction with each other. The deterministic methodology of exact classical solvers when combined with the probabilistic nature of quantum techniques may be the next step in quantum computing. A crude example of this would be to use quantum resources to explore neighboring solutions periodically dictated by a classical algorithm to optimize an arbitrary nonlinear objective function. Best of both classical and quantum domains is ensured through this approach. In the near future, along with better quality and error correction schemes for qubits and controls, there arises a need for hybrid tools capable of combining the computing power of both classical and quantum computers.

As evident from the presented case studies, quantum computers perform exceptionally better for facility-location location allocation problems. Although they perform mediocre for problems requiring discretization schemes, good feasible solutions can be obtained for problems with large solution spaces in reasonable runtimes. Sustainable design and operation of energy systems which consider economic, environmental, and social impacts are quite complex to solve [21], and a standalone quantum solution approach might not be feasible for such design problems, since the best solution is always expected in this case. However, for energy systems in the presence of uncertainty, online adaptation is required to ensure system is always operated optimally, implying the online solving of such optimization problems. Quantum computers can efficiently handle such problems, by providing good feasible solutions within reasonably shorter computational times than any classical computer. However, this does not necessarily mean that design problems will not benefit from quantum approaches. Hybrid quantum-classical methodologies can be developed for optimal design and scheduling of energy systems, which will always outperform a either of classical or standalone quantum approaches. At this stage, such algorithms need to be developed for a specific class of problems, but a more generic approach for global optimization of highly complex problems has become a necessity, and efforts need to be directed toward such technologies and algorithms.

## 7. Conclusion

In this paper, a brief introduction to the new and emerging field of quantum computing and some relevant applications to energy systems optimization are provided. The type of problems compatible with quantum hardware and quantum algorithms used for optimization are also discussed. The conducted analysis of mentioned optimization problems reveal that although an approximation by problem space discretization yields an optimal solution, its likelihood of being the best possible solution changes. Some of the challenges faced by quantum computers in terms of hardware architecture, precision and error mitigation are also stated. Although in some instances a quantum advantage may be perceived, classical algorithms explicitly customized for that particular instance could potentially outperform a quantum computer. This does not imply that quantum computers are up to no good. Quantum computing is at its earliest stages of development and still has a long way to go compared to its much matured classical counterpart. In the future, numerous applications would demand implementation of both classical and quantum resources and hence efforts need to be directed towards harnessing the power of quantum computing systems for large-scale, complex energy systems optimization problems. Problems like energy supply planning and supply chain design [22], multi-period scheduling of operations on power plants, and sustainable design and synthesis of energy systems [17] which can be further complicated by uncertainties that carry temporal correlations are examples of large-scale optimization problems for which the said quantum-classical hybrid methodologies should be implemented.

## Acknowledgements

This work was supported in part by Cornell University's David R. Atkinson Center for a Sustainable Future.

## List of References